\documentclass[lettersize,journal]{IEEEtran}
\IEEEoverridecommandlockouts
\usepackage[noadjust]{cite}
\usepackage{amsmath,amssymb,amsfonts}
\usepackage{algorithmic}
\usepackage{graphicx}
\usepackage{textcomp}
\usepackage{xcolor}
\def\BibTeX{{\rm B\kern-.05em{\sc i\kern-.025em b}\kern-.08em
    T\kern-.1667em\lower.7ex\hbox{E}\kern-.125emX}}

\begin{document}

\title{CCAT: Silicon-Platelet Feedhorns for Submillimeter Wavelengths}

\author{

 \author{Jason Austermann, James Beall, James R. Burgoyne,
 Scott Chapman, Steve Choi, Cody J. Duell, Anthony I. Huber,
 Johannes Hubmayr, Matthew A. Koc, Michael D. Niemack,
 Joel N. Ullom, Jeffrey van Lanen, Anna Vaskuri, 
 Michael Vissers, 
 and Jordan Wheeler

 \thanks{\copyright 2026 IEEE. Personal use of this material is permitted. Permission from IEEE must be obtained for all other uses, in any current or future media, including reprinting/republishing this material for advertising or promotional purposes, creating new collective works, for resale or redistribution to servers or lists, or reuse of any copyrighted component of this work in other works.}
 \thanks{Jason Austermann, James Beall, Johannes Hubmayr, Jeffrey van Lanen, Michael Vissers, and Jordan Wheeler are with the Quantum Sensors Division (QSD) at the National Institute of Standards and Technology (NIST), Boulder, CO, 80305 USA (email: jea@nist.gov)}
 \thanks{Matthew A. Koc, Joel N. Ullom, and Anna Vaskuri are with the Department of Physics at the University of Colorado, Boulder, CO, 80309, USA and the QSD at NIST, Boulder, CO, 80305 USA.}
 \thanks{James R. Burgoyne and Anthony I. Huber are with the Department of Physics and Astronomy, University of British Columbia, Vancouver, BC, Canada.}
 \thanks{Scott Chapman is with the Department of Physics and Atmospheric Science, Dalhousie University, Halifax, NS, Canada.}
 \thanks{Steve Choi is with the Department of Physics and Astronomy, University of California, Riverside, CA 92521, USA.}
  \thanks{Cody J. Duell and Machael D. Niemack are with the Department of Astronomy, Cornell University, Ithaca, NY 14853, USA}
 }
}

\maketitle

\begin{abstract}
Silicon-platelet feedhorn arrays are an established technology at millimeter wavelengths that, for some applications, can provide significant advantages over traditional direct-machined metal feedhorns. 
The Prime-Cam focal planes operating in the 350~GHz ($\sim$860~$\mu$m) and 850~GHz ($\sim$350~$\mu$m) bands are anticipated to carry the first silicon-platelet feedhorn arrays to operate fully at submillimeter wavelengths, representing a significant step forward in the application of this technology. 
In particular, the feedhorns designed for operation in the 850~GHz band represent a 3x increase in frequency compared to previously demonstrated and deployed devices of this type.
Here we present a demonstration of silicon-platelet feedhorns at these submillimeter wavelengths, including in-lab performance characterization.  We present fabrication metrology, room-temperature beam maps, and cryogenic optical efficiency measurements where the feedhorns are coupled to prototype CCAT Prime-Cam detectors.  We show that feedhorn performance measurements are well matched to simulation and compare that performance directly to traditional, direct-machined metal feedhorns.

\end{abstract}


\section{Introduction}

Silicon-platelet feedhorn arrays have previously been developed at millimeter wavelengths \cite{Britton2010,nibarger201284,simon2016,simon2018feedhorn}, primarily for cosmic microwave background (CMB)  \cite{actpol,sptpol,henderson2016_advactpol_readout,lamagna2020litebird,hubmayr2016spider} and astronomical \cite{austermann2018millimeter,duell2024ccat} experiments. 
Building feedhorn arrays using stacked and bonded silicon wafers carries several advantages over traditional direct-machined feedhorns and with no significant drawbacks related to performance.
The lithographic fabrication process results in high levels of precision and array uniformity, typically better than what is possible with traditional direct-machining of metal.
These tight tolerances become increasingly important at shorter wavelengths. 
Furthermore, the stacked platelet design allows for nearly arbitrary feedhorn profiles, including corrugated
\cite{Britton2010,nibarger201284,henning2012,hubmayr2016spider}, 
ring-loaded \cite{thornton2016}, 
or spline-profiled \cite{simon2016, simon2018feedhorn, austermann2018millimeter}
shaped walls that can be customized and optimized to match the experimental needs and priorities of any given project (e.g. bandwidth, beam symmetry, cross polarization, aperture efficiency). 
Silicon platelets also match the coefficient of thermal expansion (CTE) of the substrates used in typical detector arrays,
thus delivering material-matched advantages in the cryo-mechanical design and maintaining optical alignment from room temperature to cryogenic operation. 

The feedhorns discussed here will be deployed in Prime-Cam \cite{vavagiakis2018ccatprime} -- the first-generation science instrument currently undergoing validation for the 6-meter Fred Young Submillimeter Telescope (FYST) \cite{parshley2022_fyst} of the CCAT Collaboration. Prime-Cam consists of a 1.8-m diameter cryostat that hosts up to 7 independent instrument modules, each designed with unique observing abilities that operate in different parts of the millimeter and submillimeter spectrum.  The imaging focal plane at the heart of each module consists of large-format arrays of feedhorn-coupled microwave kinetic inductance detectors (MKIDs or lumped-element KIDs, LEKIDs) with detector and feedhorn designs that are uniquely optimized for the passband and science goals of that module. An instrument module hosts up to three arrays fabricated from 150~mm diameter silicon wafers, each comprising thousands to tens of thousands of detectors and feedhorns, depending on the wavelength band. 

To date, the shortest wavelength silicon feedhorn arrays to be integrated with detectors and deployed operate in the $\sim$280~GHz (1.1~mm) band \cite{hubmayr2016spider,austermann2018millimeter,duell2024ccat}.
Here we push the technology into the submillimeter and present in-lab characterization of silicon-platelet feedhorns for operation in bands centered at either 350~GHz or 850~GHz. 
We note that concurrent development of single-pixel prototype submillimeter wave silicon-platelet feedhorns for the LiteBIRD experiment,
in bands up to 402~GHz, are being published \cite{stevenson_litebird_temp2025} in tandem with this work.
Here we describe a new design approach and updated fabrication methods that are generally applicable to this technology.
We include measurements of full 150~mm diameter production arrays destined to be deployed in Prime-Cam.
We present room-temperature beam maps of the 350~GHz band feedhorns and cryogenic performance measurements of the 850~GHz devices when coupled to Prime-Cam style detectors.
Where measured, we find that the feedhorns produce beams that are well matched to simulations, exhibit high levels of uniformity across large-scale arrays, and perform with high optical efficiency (low loss).

\section{Feedhorn Design}
\label{sec:design}

\begin{figure}[tb]
\centerline{\includegraphics[width=\columnwidth]{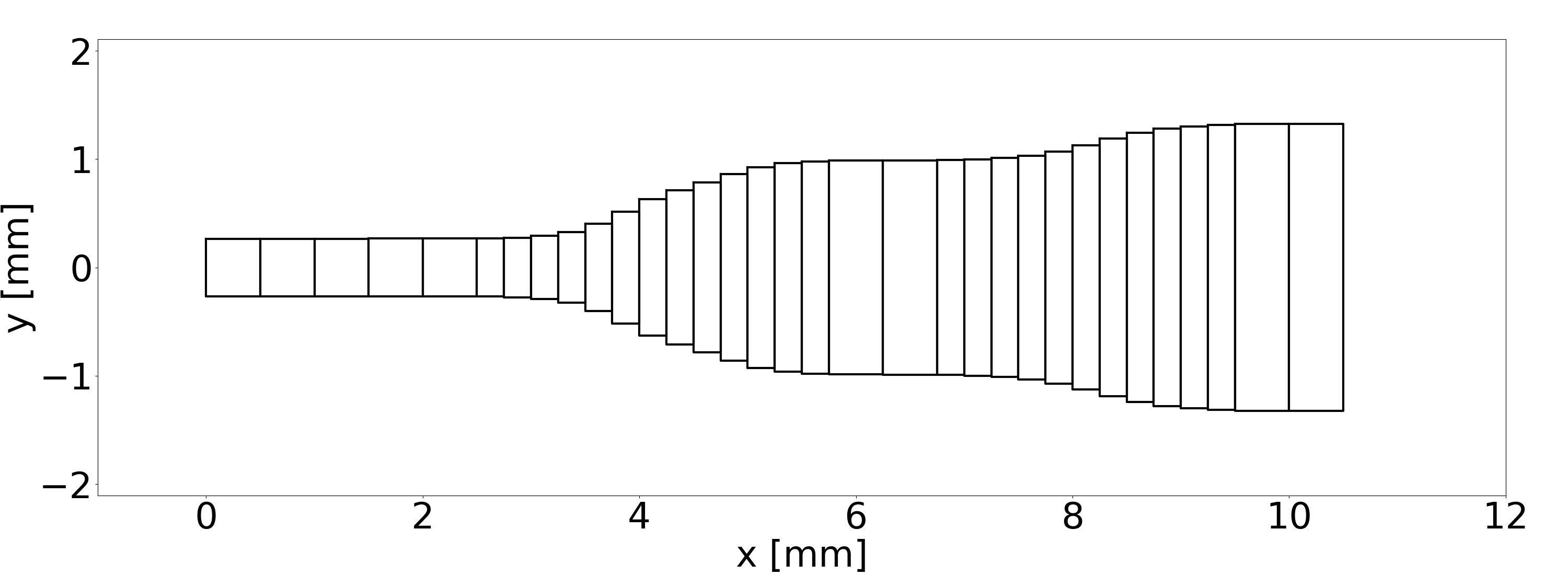}}
\caption{
The CCAT 350~GHz feedhorn design optimized for performance between 330~GHz and 370~GHz. 
The feedhorn shape is built from 33 individual platelets, each with a unique hole size and a wafer thicknesses of either 250 or 500~$\mu$m. The smaller step size is required where the profile is rapidly changing diameter and approximates a smooth-walled feedhorn. 
}
\label{fig:350design}
\end{figure}

Several different classes of feedhorn profiles and array sizes have been developed using silicon-platelet technology, with each design tuned to the needs of the end-use experiment. 
For the CCAT submillimeter bands, we have chosen to explore spline-profiled feedhorns for their combination of manufacturability and their ability to produce beams with high symmetry, low cross-polarization, and high beam coupling efficiency through the optical cold stop (i.e. minimizing spillover) \cite{simon2016}.

A new process was developed for optimizing spline-like feedhorn profiles of this type.  
Previous generations of silicon platelet feedhorns were designed as described in \cite{simon2016}, where  parallel Markov chain Monte Carlo (MCMC) instances produced numerous optimizations from randomly generated seed profiles.  
In that method, profiles were forced to monotonically increase in size from waveguide to sky-side aperture as to also be manufacturable via traditional direct machining.  
Profiles were optimized according to experimental priorities, e.g. beam symmetry in \cite{simon2016} or a more complex set of metrics in the penalty function in \cite{barron2022}.
While the goal of an optimized feedhorn profile remains the same, we present an all new design package that was created and built as to more efficiently explore a potentially larger dynamic range of fully parameterized feedhorn profiles that do not require monotonically increasing profiles (platelet based feedhorns can have nearly arbitrary shape).
This optimization process prioritizes flexibility in the feedhorn parameterization and the weighted performance metrics as to tune the optimization process to a particular experiment, and its systematic priorities, at the onset of the process that we describe as follows.

The design process begins with user-defined fixed aperture sizes at the waveguide and sky-side ends of the feedhorn.  
The waveguide diameter is typically defined to provide a high-pass filter for the desired frequency band, while the sky-side aperture is set by the pixel pitch of the detector array minus the $\sim100~\mu$m interstitial silicon remaining between feedhorns for structural integrity (this is comparable to the $\ge$100~$\mu$m needed for aluminum feedhorns \cite{simon2018feedhorn}). 
For the feedhorns presented here, the profile shape was parameterized as a series of logistic function 'S'-shape curves that smoothly connect the user-defined aperture sizes at the waveguide and sky-side ends of the feedhorn. 
These flaring sections have three free parameters: location along the feedhorn, amplitude, and slope.  The amplitude of one flare is degenerate with the user-defined aperture sizes and amplitudes of the other flares such that the number of free parameters is $3N-1$, where $N$ is the number of parameterized flares.

\begin{figure}[tb]
\centerline{\includegraphics[width=\columnwidth]{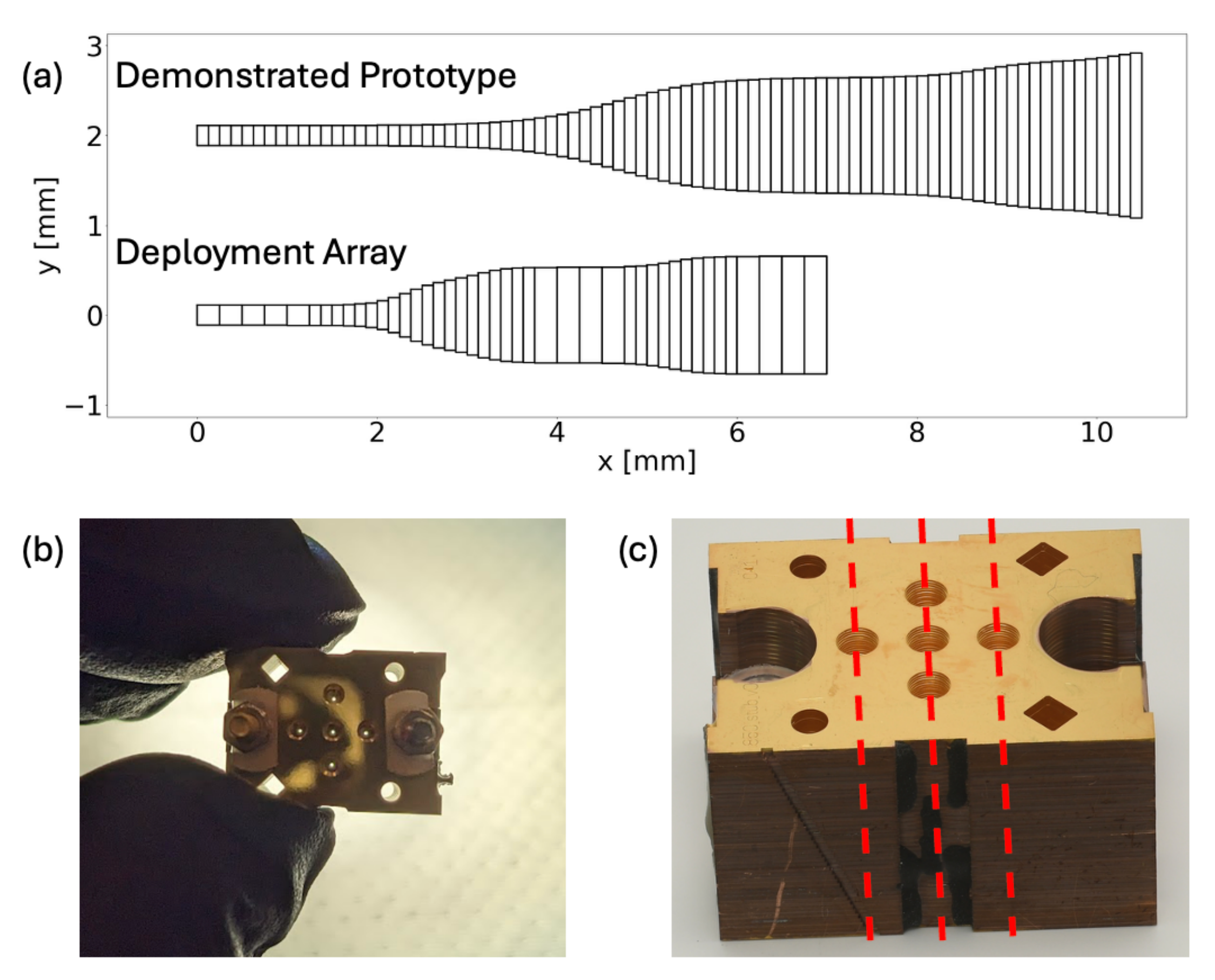}}
\caption{
The CCAT 850~GHz feedhorn designs and initial prototypes. 
(a) The as-built prototype profile (top) and the deployment array profile (bottom), each optimized for performance between 800 and 900~GHz. 
The prototype design is a truncated version of a 2~mm pitch design 
that was fabricable from the number of platelets produced from a single 150~mm diameter wafer. 
The prototype design has an extended length waveguide to strenuously test our ability to electroplate the full length of a waveguide of that diameter. 
The deployment profile is a 1.4~mm pitch (1.31~mm aperture) feedhorn with a shorter waveguide and profile.
(b) One of two 5-pixel prototypes fabricated for operation in the 850~GHz band.
(c) The same prototype with dashed lines depicting the wire-dicing axes for the cross sectioned part shown in Fig.~\ref{fig:850inspect}.
}
\label{fig:850design}
\end{figure}

The optimal feedhorn shape for a given experiment is determined using a differential evolution minimization algorithm that attempts to avoid local minima and find the global optimization based on four performance metrics: beam coupling, beam symmetry, cross-polarization, and reflection (S11). 
We construct the penalty function to be minimized as,
\begin{equation} \label{eq:}
p \equiv 1 - \bar{\eta}_{\rm{beam}} + w_{\rm{1}}\bar{P}_{\rm{sym}} + w_{\rm{2}}\bar{P}_{\rm{x}} + w_{\rm{3}}\bar{P}_{\rm{S11}} + w_{\rm{4}}\sigma(P_{\rm{sym}}),
\end{equation}
where $\bar{\eta}_{\rm{beam}}$ is the band-averaged beam coupling efficiency, $\bar{P}_{\rm{sym}}$ is the band-averaged difference in power through the cold stop between the E- and H-plane beams relative to total power,  $\bar{P}_{\rm{x}}$ is the band-averaged cross-polarization as a fraction of total power, 
$\bar{P}_{\rm{S11}}$ is the band-averaged fraction of input power reflected by the feedhorn,
and $\sigma(P_{\rm{sym}})$ is the standard deviation of the fractional power asymmetry across the band. 
Weights, $w_{\rm{n}}$, are applied to each of these performance metrics and are user-defined as to match the performance priorities of the experiment.
Performance metrics are calculated for every iteration using
beam patterns generated using the commercial mode-matching software package RTCC\footnote{http://www.smtconsultancies.co.uk/products/rtcc/rtcc.php.}.

\begin{figure*}[tb]
    \centering
    \begin{minipage}{0.45\textwidth}
        \centering
        \includegraphics[width=\textwidth]{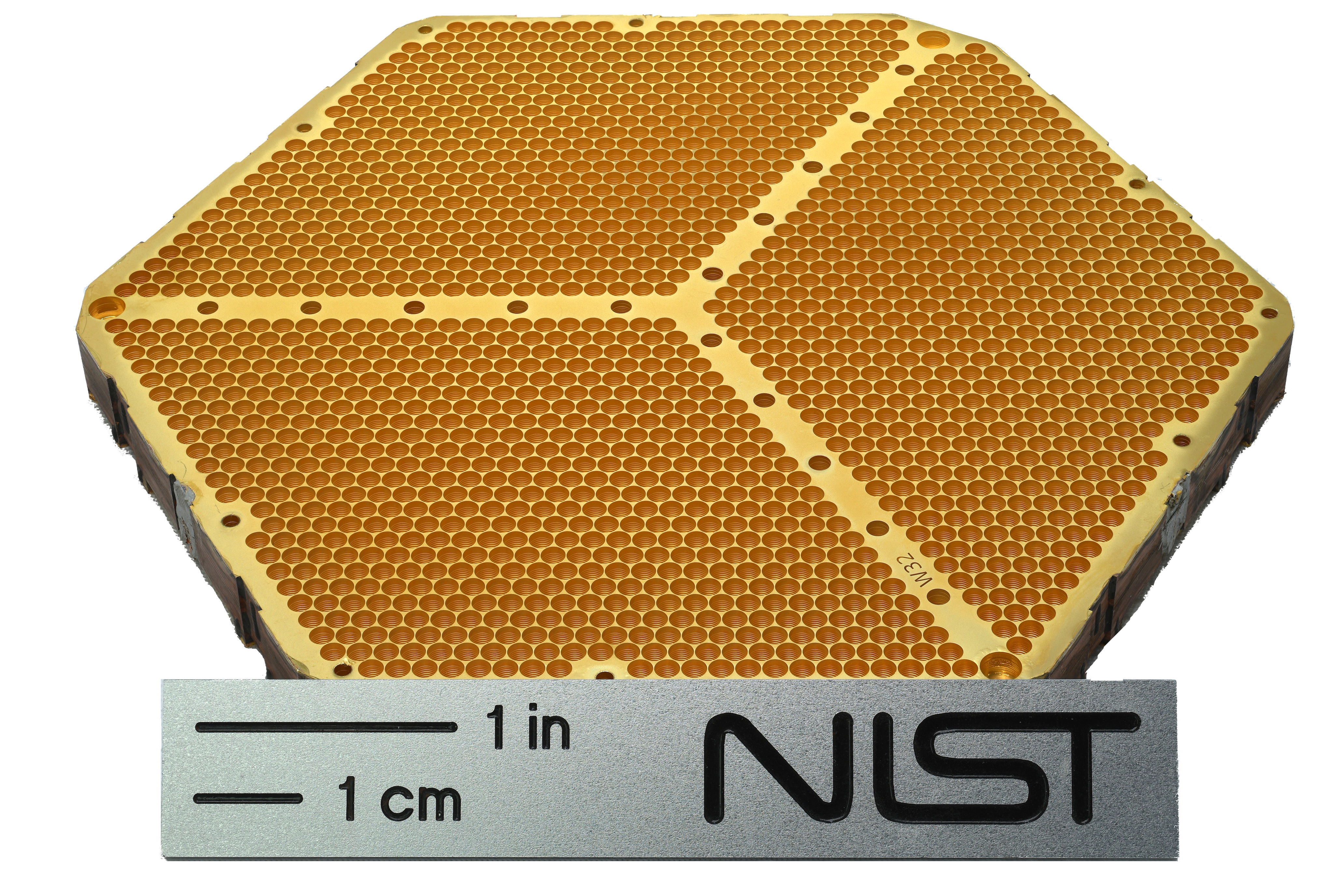}
    \end{minipage}\hfill
    \begin{minipage}{0.55\textwidth}
        \centering
        \raisebox{-0pt}{\includegraphics[width=0.82\textwidth]{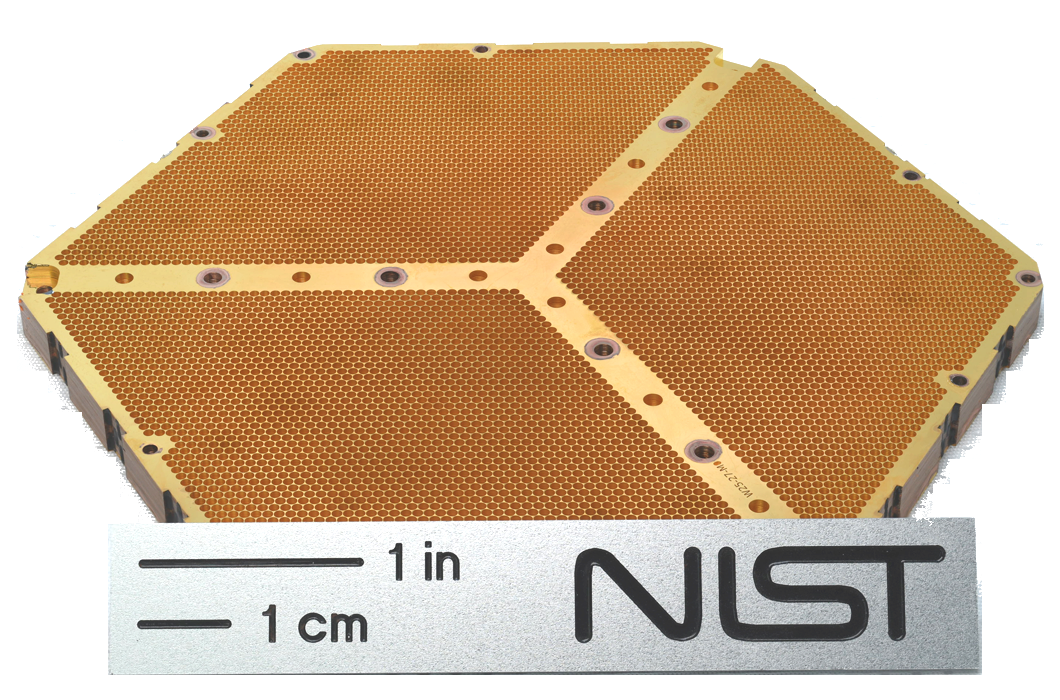}}
    \end{minipage}
    \caption{
A full production CCAT 350~GHz feedhorn array (left) and 850~GHz array (right).
The arrays are produced using multiple etched, sputtered, stacked, and aligned 150~mm diameter silicon wafers.  
The stack is electroplated as a single unit to produce a smooth, continuous conductive surface inside the feedhorn.
The array footprints are $\sim$140~mm from corner to corner and comprise 1,707 and 6,288  feedhorns for the 350~GHz and 850~GHz arrays, respectively.  
Each frequency band instrument module of Prime-Cam will contain 3 such arrays tiled into one focal plane.  }
    \label{fig:fabricated_arrays}
\end{figure*}

The resulting optimized feedhorn profile largely depends on the chosen performance metric weights.  For projects that prioritize high beam symmetry and low cross-polarization, we typically find the optimized feedhorn shape to have two or three distinct flaring sections, 
each separated by relatively flat sections in the profile (e.g. Fig.\ref{fig:350design} and Fig.\ref{fig:850design}). This is qualitatively similar to that found by similar optimizations in \cite{simon2016,simon2018feedhorn,stevenson_litebird_temp2025}.
For the feedhorn optimizations discussed here, the parameterized profile was allowed to have four flaring sections, however, the optimal shapes were found to have two flaring regions that were
produced by preferred parameters merging the four allowed features into two more complex structures. 
We explored allowing up to five independent parameterized features, but the results were similar with performance metrics equal at the sub-percent level. 

These platelet feedhorns have an inherently stepped profile due to the approximately straight-wall holes of each layer (see Sec.~\ref{sec:fab}). 
Through simulation, we find that the maximum step size (i.e. thickness of each silicon step) in the rapidly changing profile regions should be $\lesssim  0.5\lambda$, above which frequency-dependent resonant features and reflection spectra begin to appear in the performance metrics.
In these regions, we choose to use step sizes of 250~$\mu$m and 125~$\mu$m for the 350~GHz ($\lambda \sim$ 860~$\mu$m) and 850~GHz ($\lambda \sim$ 350~$\mu$m) bands, respectively. We find that using simulated step sizes smaller than these values result in no significant change in performance or beam patterns in their respective frequency bands.
During the optimization process, these discrete step sizes are used for every layer in the profiles. 
After an optimized profile is selected, some of the steps in the flat areas of the profile are merged into a smaller number of long steps to create an equivalent profile with negligible effect on performance while reducing the number of wafers to fabricate.

The total length of the feedhorn is fixed in each optimization run in order to reduce the parameter space explored.  
Instead, several different lengths were individually optimized for each band. In general, we find that once a certain minimum length is reached, the resulting optimizations are similar in shape, with any added length going into an extended waveguide and the performance is equivalent at the sub-percent level. 
The final feedhorn designs were chosen to be near the minimum length that meets this threshold and are shown in Figs.~\ref{fig:350design}~\&~\ref{fig:850design} for the 350 and 850~GHz bands, respectively.  

The 350~GHz experiment optics and pixel pitch of 2.75~mm were set early in development and was the only pitch explored for optimization.
The designed waveguide diameter of 0.532~mm provides a high-pass cutoff of $\sim$330~GHz. 
The feedhorn and waveguide total length is 10.5~mm and has a sky-side aperture diameter of 2.65~mm.

Development of the 850~GHz camera explored multiple optical configurations, resulting in optimized feedhorn designs for both 2.0~mm and 1.4~mm pixel pitch options.
A truncated version of the 2.0~mm pitch feedhorn, with an extended-length waveguide, was chosen for the initial prototype demonstration.
Later, the 1.4~mm pitch design (1.31~mm sky-side aperture) was chosen for the final, full-arrays as this pixel density is projected to result in higher mapping speeds.  
This design has a 0.225~mm diameter waveguide and a total length of 7~mm.

\section{Fabrication Methods and Metrology}
\label{sec:fab}
The submillimeter silicon-platelet feedhorns were fabricated using processes that have evolved from those  described for previous millimeter-wave feedhorns (e.g. \cite{Britton2010,nibarger201284,simon2016,austermann2018millimeter}).
Each silicon platelet layer in the stack has holes defined using photolithography and is micromachined using deep reactive ion etching (DRIE) through the Bosch process. 
Each platelet individually receives seed layer metallization on both sides in the form of 200~nm of sputtered Ti followed by 
$\sim$~1~$\mu$m of Cu. 
The full set of platelets are then stacked and aligned.
Earlier generations of silicon-platelet feedhorns used stainless steel pins for alignment (e.g. \cite{Britton2010,nibarger201284}), however, this was found to lead to racking (tilting) of the stack relative to the optical axis. 
Over the years, an improved approach was developed that aligns the edges of the stack against three precision granite blocks, each with sub-micron surface flatness.  
While monitoring under a microscope, each platelet has two of its photolithographically-defined flat edges carefully pushed and tapped into place against the reference blocks.  
The aligned stacks are held together and in alignment by tightening pre-installed screws 
that are later removed when full fabrication is complete. Small strips of Stycast 2850FT epoxy are applied in predefined
notches along the sides of the stack and serve to lock in and maintain stack alignment. 
Finally, the fully aligned stack is electroplated\footnote{All NIST-produced silicon-platelet feedhorns have been commercially electroplated at Custom Microwave Inc., now Vitesse Systems, vitessesys.com.}
with a target of $\sim$3~$\mu$m of Cu for its gap-filling properties, followed by $\sim$3~$\mu$m of Au for continuous electrical conductivity down the feedhorn and waveguide throat. 
The bare etched holes of each layer are intentionally over-sized by $6\mu$m in radius relative to the optimized design to account for this metallization thickness. 
Examples of completed arrays are shown in Fig.~\ref{fig:fabricated_arrays}.

\begin{table*}[t] 
\centering
\caption{
Platelet hole radii before electroplating as measured from a 3D reconstruction of the platlet surface using a confocal laser. 
Each measurement is the average of 5 holes in the same region of the wafer (center, mid-radii, edge) and the 1$\sigma$ standard deviation of that set.
Wafer numbers increase from the waveguide aperture (wafer 0) to the sky-side aperture (wafer 27). Wafer 13-L is a measurement of the large hole side of a double-etched, two-step wafer. All others are single etch wafers. 
A lower magnification objective lens is used for the larger hole sizes, leading to larger uncertainty in the measurements.
}
\label{tab:hole_radius}
\begin{tabular}{ |c|c|c|c|c|c|  }
\hline
Wafer \# & Design Radius [$\mu$m] &  Center of Array [$\mu$m] &  Middle of Array [$\mu$m]  & Edge of Array [$\mu$m] & Design-Meas. Diff. Avg. [$\mu$m]\\
 \hline
 00   & 118.50 & 123.21 $\pm$ 0.05& 123.31 $\pm$ 0.02 & 123.3 $\pm$ 0.07 & 4.77 $\pm$ 0.07 \\
 05   & 119.70 & 124.16 $\pm$ 0.07& 123.68 $\pm$ 0.07 & 124.19 $\pm$ 0.05 & 4.31 $\pm$ 0.25 \\
 13-L   & 520.80 & 526.4 $\pm$ 1.2 & 525.10 $\pm$ 0.04 & 525.00 $\pm$ 0.75 & 4.68 $\pm$ 1.07 \\
 27   & 661.00 & 663.10 $\pm$ 0.25 & 663.53 $\pm$ 0.49 & 664.2 $\pm$ 0.47 & 2.61 $\pm$ 0.63 \\
 \hline
\end{tabular}
\end{table*}

The alignment of stacked silicon using the granite blocks technique has been spot checked for various stacked-silicon structures at NIST. 
We find that the layer-to-layer alignment is typically better than 5~$\mu$m. 
A simulated tolerance study on multiple platelet feedhorn designs has previously shown that a normal distribution of layer-to-layer misalignment with $\sigma=$3~$\mu$m, including all random deviations significantly larger, 
has negligible affect on the beam properties up to at least 450~GHz \cite{lamagna2020litebird}.  In that study, even the largest induced effects on the beam resulted in negligible performance changes and those outliers were driven by the largest random misalignment realizations in the distribution.  
In practice, the largest deviations will likely be identified during the under-microscope alignment process and improved before permanently setting the alignment. 
Furthermore, the largest effects are on the beam sidelobes, which are terminated at the cold stop of the instrument optics and play little to no role in sky measurements. 
Experimentally, no degradation in performance has yet been identified as resulting from misalignment of the silicon platelet feedhorns assembled in this manner, including the highest frequency arrays deployed to date at 280~GHz \cite{hubmayr2016spider,austermann2018millimeter} and the 350~GHz arrays presented here.  

The 850~GHz feedhorn and waveguide present the tightest required tolerances yet attempted for a feedhorn of this type. 
To understand the tolerance on alignment, we performed a series of electromagnetic simulations on the first 2~mm of the feedhorn which includes the full waveguide and waveguide coupling structure. 
This section of the feedhorn is the most sensitive to misalignment with respect to reflection/transmission and defines the high-pass filter quality of the waveguide. 
For this study, we simulated 1000 realizations where each platelet is randomly misaligned using a normal distribution with $\sigma=$~3~$\mu$m.
All simulated realizations resulted in performance within tolerance of the experiment and the vast majority experienced negligible affects.
We find that, relative to the perfectly aligned case, $>$ 99.8\% of the realizations retain $\ge$99\% transmission of the primary TE11 waveguide mode.   
Induced cross polarization is less than 0.1\% in 95\% of the realizations.
In addition, such a population of misalignments results in a negligible average shift of the waveguide cutoff frequency of $\delta f=$ 0.25~GHz with a standard deviation of 0.33~GHz -- a negligible amount that is equivalent to a less than 0.1~$\mu$m change in the waveguide radius and much smaller than the uncertainty in metallization thickness which is discussed later in this section. 

All feedhorn platelets were fabricated from 150~mm diameter silicon wafers.
For both the 350~GHz and 850~GHz designs, 5-pixel sub-array prototypes (e.g. Fig.~\ref{fig:850design}) were fabricated and tested before moving on to production of the full-scale arrays seen in Fig.~\ref{fig:fabricated_arrays}.
While only one platelet is produced per wafer for full-scale feedhorn arrays, 42 platelets are yielded from a single wafer for the small 5-pixel sub-arrays, thus enabling rapid prototyping. 
Wafer thicknesses are tuned to the specific designs of Sec.~\ref{sec:design}.
The 350~GHz feedhorn profile is produced from a combination of 250~$\mu$m and 500~$\mu$m thick platelets with single-etch through holes.  
The 850~GHz profile is created using only 250~$\mu$m thick wafers, but with many of the layers consisting of a 2-tiered etch\cite{Britton2010} to produce two separate 125~$\mu$m steps with unique diameters. 
All platelet holes are designed to have straight sidewalls 
using a modified Bosch DRIE processes and an SPTS Rapier.
Scanning electron microscope images of cleaved platelet layers show typical sidewall DRIE tapers of $\sim$0.5~deg -- a significant improvement compared to past fabrication methods and equipment that resulted in $\sim$2.5~deg slopes, which can significantly affect waveguide performance at these higher frequencies\cite{vaskuri2025}. 

\begin{figure}[t]
\centerline{\includegraphics[width=\columnwidth]{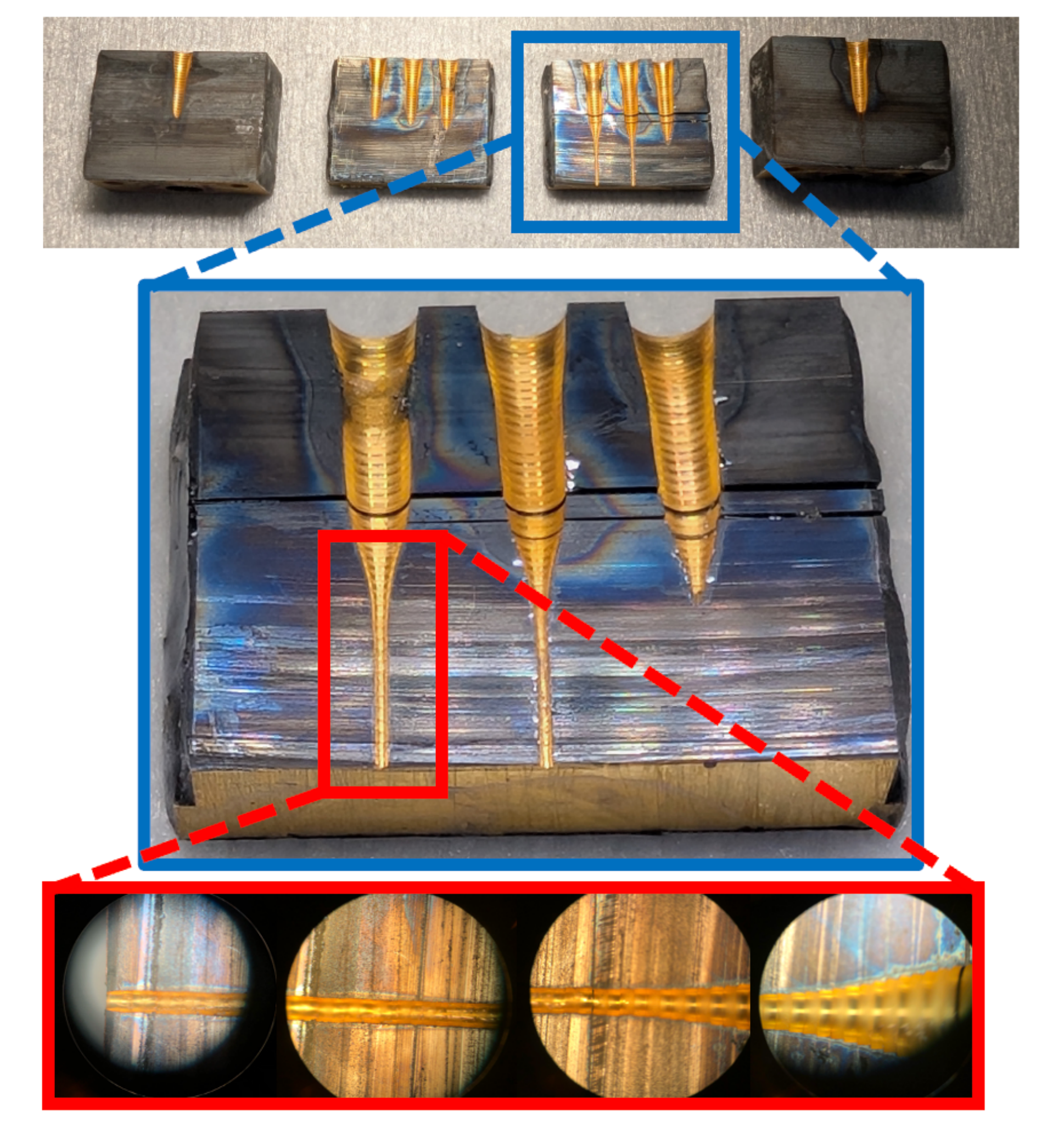}}
\caption{
A wire-saw diced 850~GHz prototype feedhorn sub-array.
All exposed waveguide and feedhorn surfaces were found to be smoothly and uniformly coated with the electroplated gold, including through the extended-length waveguide without any clogging or excess accumulation. 
The wire saw kerf is similar to the waveguide diameter, therefore the waveguide only survives on one of the two halves of an offset cut. 
A slight angle in the cut axis resulted in one waveguide being completely removed on the highlighted piece.
Flexible translucent epoxy was used to fill and protect each feedhorn during dicing and later removed. 
Residual epoxy remains in some areas, notably in the upper left of the highlighted feedhorn and a filled waveguide of the rightmost part in the upper image.   
The relatively violent sawing process resulted in the separation of several platelet layers, but this has not been an issue under normal use and handling (see text).
}
\label{fig:850inspect}
\end{figure}

The waveguide diameter of the 850~GHz feedhorn is over 3x smaller than any previously produced silicon-platelet feedhorn and $\sim$10x smaller in area.  
This posed a potential challenge to our fabrication procedures and, in particular, whether the metallization processes would successfully and sufficiently coat the critical surfaces continuously and without creating unintended structures or blockages in the small waveguide. 
To test fabricability of the 850~GHz feedhorns, two identical 5-pixel prototype sub-arrays were produced (Fig.~\ref{fig:850design}) using the DRIE process.
These prototype feedhorn profiles were given an extended length waveguide to present a more challenging metallization process than would be faced for the shorter production array feedhorns. 

Inspection of the completed 850~GHz prototypes found that their waveguide apertures have diameters consistent with expectation and no visible indications of clogging or accumulation of metals at the aperture. 
One of these sub-arrays was cross-sectioned using a wire saw to allow inspection of the full throat of the feedhorns.
Three cuts were made along the axes shown in Fig.~\ref{fig:850design} and the metallization was inspected under a microscope. 
Images of the diced feedhorns are shown in Fig.~\ref{fig:850inspect}.
We find the metallization of all feedhorns to appear evenly distributed, continuous, and with no visible signs of excess buildup or blockages.  
The rough surface of the cut plane did not allow for a reliable estimate of the achieved metallization thickness in cross-section, but there were no visible indications that the waveguide plating was of a significantly different thickness than the larger portions of the feedhorn.
The relatively violent process of dicing the arrays caused some platelet layers to separate, most notably near the middle of the feedhorn profile 
where we lacked epoxy on the sides to hold the layers together.
Production arrays routinely have epoxy connecting all platelets at multiple locations around the periphery and no platelet separation has been observed in any silicon-platelet feedhorn under normal use and handling, including in all deployed arrays \cite{grace2014,henderson2016,wilson2020toltec,bergman2018-280,austermann2012sptpol}.

\begin{figure}[t]
\centerline{\includegraphics[width=\columnwidth]{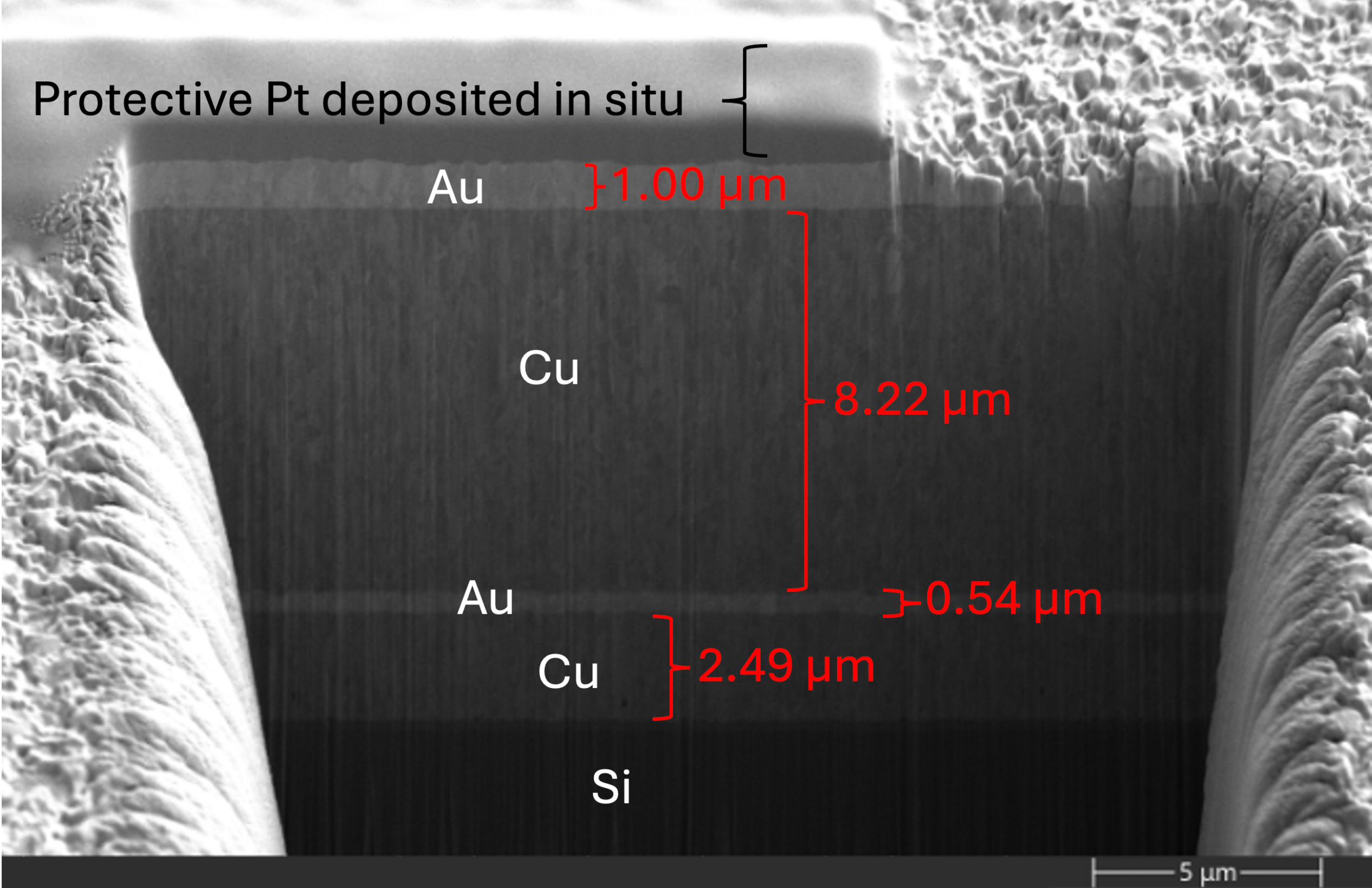}}
\caption{
A focused ion beam scanning electron microscopy (FIB-SEM) cross-sectional image of the electroplated surface of a 350~GHz feedhorn prototype.  
A layer of Pt is deposited in the FIB milling area to protect the top gold surface, which is otherwise roughened in the process as seen outside the deposited Pt area.  
This prototype underwent two rounds of successive Cu/Au plating to fine tune the final waveguide diameter via total metal thickness.
The bottom Cu layer is a combination of the sputtered seed layer Cu ($\sim$1~$\mu$m) and electroplated Cu.
}
\label{fig:350fib}
\end{figure}

The accuracy and uniformity of our platelet hole sizes was measured using confocal laser optical profilometry. This is a time-consuming process, therefore, these values were spot checked at multiple locations on a subset of platelets from the full 850~GHz array.  
A summary of these measurements is presented in Tab.~\ref{tab:hole_radius}.
While the arrays appear to be highly uniform from center to edge, there does appear to be a bias in the average hole size radius of approximately +$4\mu$m. Simulations show that hole size biases of this size are insignificant to feedhorn performance and beam shape, but do affect the cutoff frequency of the high-pass filter provided by the waveguide.
For future fabrications, we expect it to be straightforward to modify our process as to correct for this bias.

The metallization thickness also affects the waveguide radius and cutoff frequency. 
We have estimated the metal thickness of the 350~GHz prototypes and arrays through optical microscopy and the total added mass of electroplating.  We've also measured the high-pass cutoff of the 350~GHz band waveguides using a room-temperature millimeter-wave vector network analyzer (VNA), as well as cryogenically when coupled with detectors using a Fourier transform spectrometer (FTS). 
All measurements were consistent with the devices receiving an average realized metallization thickness of $\sim$40\% of our combined Cu+Au goal of 6~$\mu$m.
The thinner electroplating has been directly confirmed on the surface of one of the prototype 350~GHz feedhorn blocks using focused ion beam scanning electron microscopy (FIB-SEM), as can be seen in the first round of deposited layers of Fig.~\ref{fig:350fib}.

The thinner metallization does not impact the low-loss transmission of the feedhorn and waveguide as just $\sim$1~$\mu$m of Cu+Au is roughly 10 skin depths at room temperature.
However, when combined with the over-etch bias and slight hole taper, this results in a waveguide cutoff frequency of $\sim$320~GHz (design value 330~GHz) and allows transmission of a bright and unwanted atmospheric water-vapor line at 325~GHz.
To correct for this, a second round of electroplating is applied to all 350~GHz arrays as to fine tune the waveguide cutoff to $\sim330$~GHz (Fig.~\ref{fig:350fib}). 
The electroplating used is a timed process that will be re-calibrated for future arrays.
The 850~GHz arrays are relatively unaffected by this thinner electroplating as the experiment does not rely on the waveguide to provide a high-pass filter for band definition and instead uses free-space metal-mesh filters\cite{ade} to define both edges of the band.  This allowed the 850~GHz feedhorns to have intentionally over-sized waveguides that improve in-band detector coupling and present a larger cross-section to the metallization process.

The focal planes of the final instruments will each consist of 3 full-scale arrays. 
All three 350~GHz feedhorn arrays have been manufactured, as has the first of the 850~GHz arrays (Fig.~\ref{fig:fabricated_arrays}).
Two of the 350~GHz arrays had their platelets micromachined using the DRIE process described previously, while the third set of platelets was produced using laser micromachining.  
Laser micromachining of silicon platelets for use in feedhorns has previously been explored for millimeter-wave use \cite{simon2018feedhorn}, although that work did not lead to a fully realized feedhorn.
Here, we have fabricated a full 350~GHz array primarily from laser micromachined parts in a research partnership with the Inertial Fusion Technology (IFT) Division at General Atomics.
Measurement and analysis of that array is ongoing and will be the subject of a future publication.

\begin{figure}[tb]
\centerline{\includegraphics[width=\columnwidth]{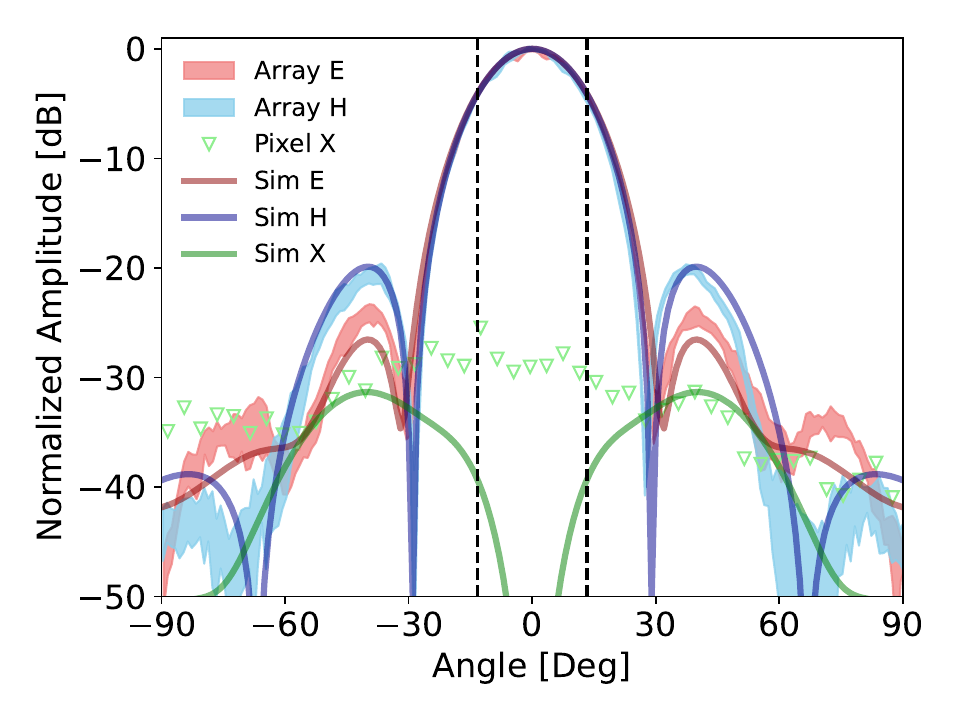}}
\caption{
Beam profile measurements (filled regions and data points) of the first full-production 350~GHz band feedhorn array at 336~GHz as compared to simulation (solid lines).
The distribution of H- and E-plane beams are shown as the +/- 1$\sigma$ range of six measurement locations spanning the array.
This range is comparable to the systematic variance due to optical alignment of each measurement and thus represents an upper limit to the physical beam variation expected across the array.
Cross-polarization (X) is a low-signal measurement and particularly susceptible to systematics in experimental alignment. 
Plotted (green) is the average X measured in three alignment attempts of the same pixel, which we interpret as an upper limit that is consistent with simulated expectations and well below CCAT requirements.
Vertical dashed lines represent the average opening angle to the cold stop in the CCAT 350~GHz optics design, inside of which is the portion of the beam that couples to the sky.  
}
\label{fig:350meas}
\end{figure}

\section{Measurements}
\label{sec:meas}

The beams produced by the 350~GHz feedhorns are measured using a room-temperature millimeter-wave VNA 
setup similar to that described in \cite{simon2016}.  
Here, the feedhorn under test (DUT) is coupled to a WR3.4 waveguide and used in broadcast mode as the transmitter in the setup.  
The WR3.4 configuration has a maximum operational frequency of 336~GHz, so we only measure the lower frequencies of the 350~GHz band.
A commercial diagonal feedhorn is used as the receiver and placed in the far-field of the DUT for all E and H plane measurements, while a prototype 350~GHz feedhorn is used for cross-polarization measurements.
The beam is measured over a range of angles by rotating the receiver about the approximate phase center of the DUT at a constant distance while always pointing at the DUT.
The receiver is kept at a sufficient distance and has a large enough beam to effectively see the DUT as a point source that is kept in the center of the receiver's beam; therefore, the details of the receiver's beam pattern do not impact the results. 

\begin{figure}[t]
\centerline{\includegraphics[width=\columnwidth]{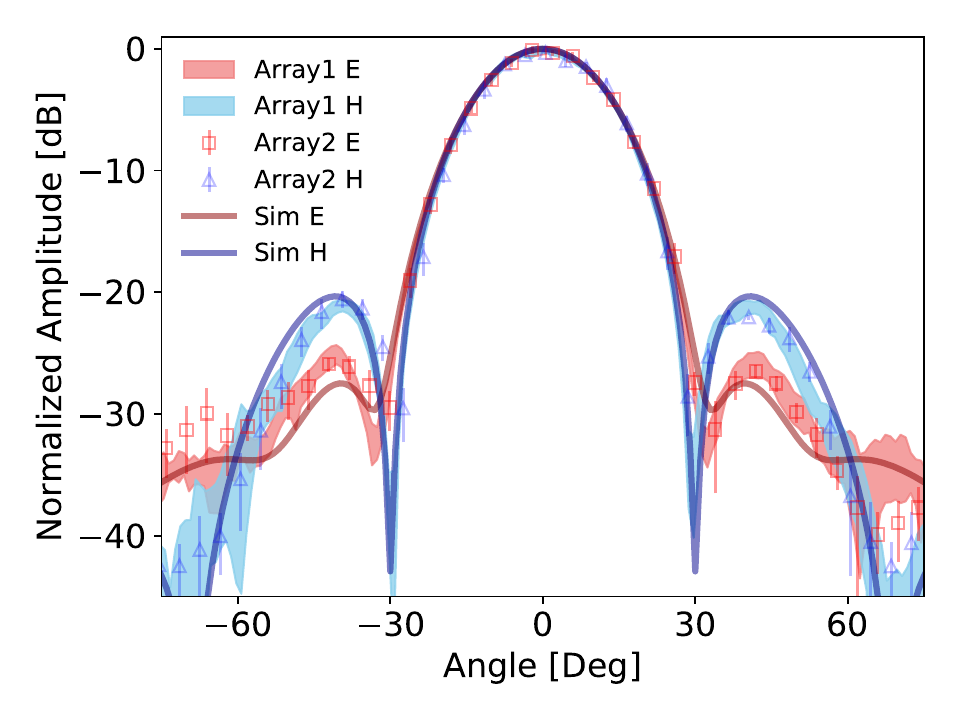}}
\caption{ H- and E-plane beam measurements at 330~GHz for two separate CCAT 350 GHz production arrays. 
Shaded regions and data points with error bars represent the average and +/-1$\sigma$ standard deviation from measurements of multiple feedhorns spanning the full array. Six feedhorns were measured for Array 1 and five for Array 2.  All beam measurements are sampled at 1~deg steps, but for plot clarity the Array 2 data is averaged at 4~deg resolution. 
The simulated beam patterns are depicted as solid lines.
}
\label{fig:350meas2}
\end{figure}

We measured the beam profile at various feedhorn locations across the first two 350~GHz production arrays.
Measurements at 336~GHz are shown in Fig.~\ref{fig:350meas} for the first production array, while the two arrays are directly compared at 330~GHz in Fig.~\ref{fig:350meas2}.
We find that the measured beams are highly uniform across the arrays are broadly consistent with simulations and previously fabricated prototype feedhorns, with minor differences within the range of expected systematic uncertainties in the alignment of the measurement system.
All array measurements were made while maintaining the same system alignment, however, the act of remounting the array for each feedhorn location likely resulted in small alignment changes between measurements, as evidenced by small changes observed in measurements of successive mountings of the same pixel.  Therefore, the observed spread in beam measurements provides an upper limit to the inherent variation of beam properties across the array.

\begin{figure}[t]
\centerline{\includegraphics[width=\columnwidth]{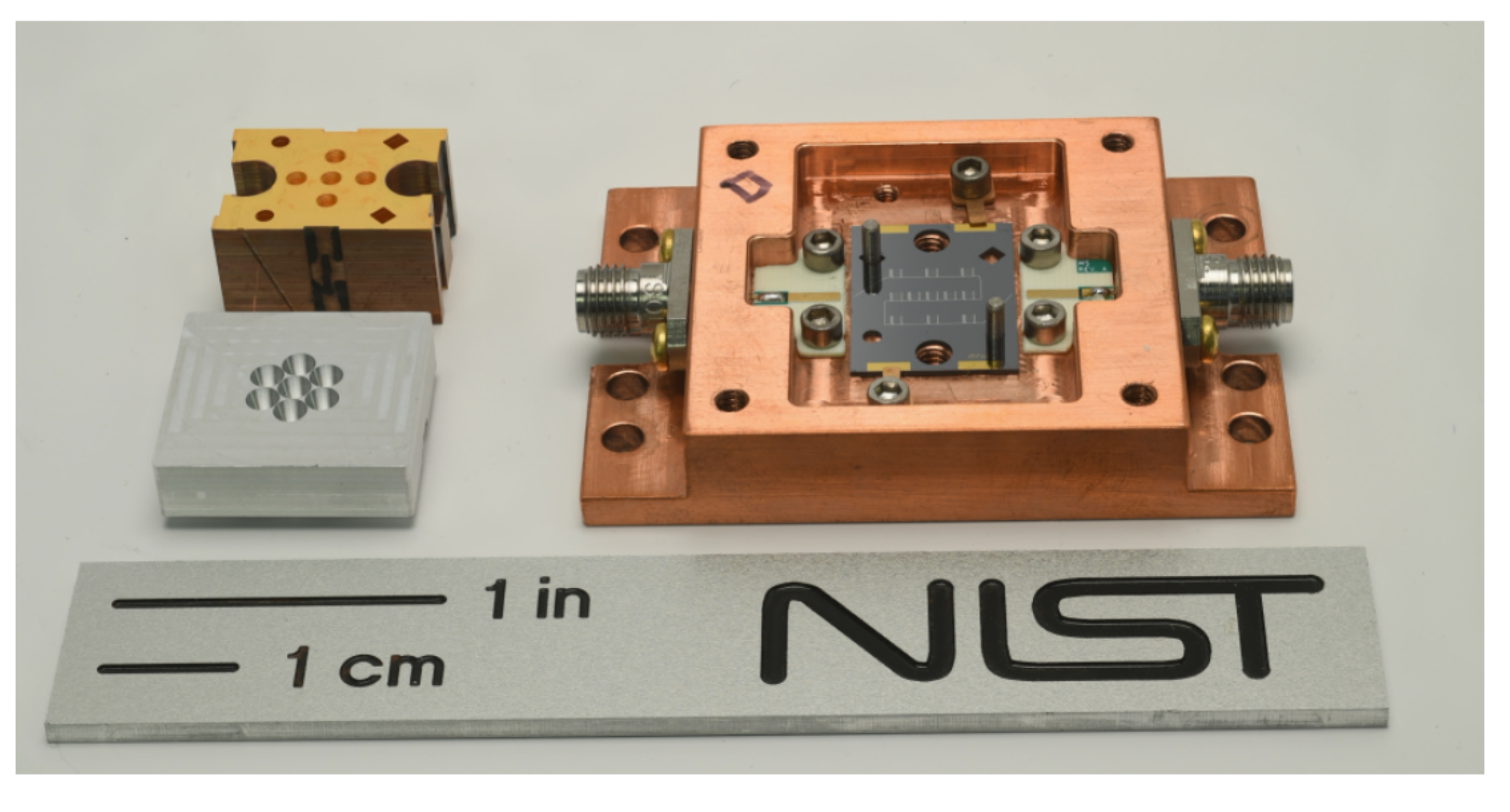}}
\caption{The CCAT 850~GHz prototype measurement setup.  
Prototype 850~GHz band detectors are mounted in a copper box on the right.  The prototype silicon-platelet feedhorn sub-array (upper left) and reference direct-machined aluminum feedhorns (lower left) are mounted atop the detectors and aligned via alignment pins and visual inspection under a microscope.
Four metal-mesh $\sim$1~THz low-pass filters (not shown) are used between the feedhorns and blackbody source in the measurements of Fig~\ref{fig:850meas}
}
\label{fig:850setup}
\end{figure}

\begin{figure}[tb]
\centerline{\includegraphics[width=\columnwidth]{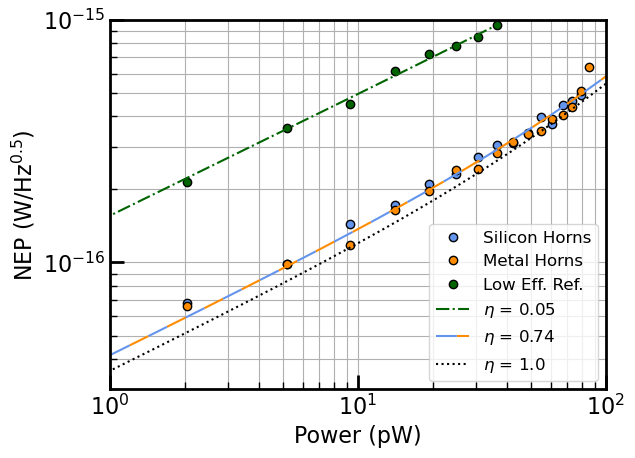}}
\caption{Noise equivalent power (NEP) and fitted optical efficiency measurements of prototype CCAT 850~GHz detectors when alternatively using the prototype silicon-platelet feedhorns (blue) and aluminum direct-machined feedhorns (orange) to couple detectors to a variable-temperature cryogenic blackbody.
Data points represent NEP measurements while lines are the expectations for various optical efficiency values.
See text for details.
As a reference, we compare to a low efficiency measurement (green) where a section of waveguide with incomplete metallization was placed between the detector and feedhorn structure. 
}
\label{fig:850meas}
\end{figure}

We do not currently have access to components to operate the room-temperature beam mapping equipment within the 850~GHz band.  
However, the prototype feedhorns were cryogenically tested as the receiver using prototype kinetic inductance detectors for the CCAT 850~GHz camera \cite{Chapman2023}. 
We directly compare the optical efficiency of the prototype silicon-platelet feedhorns to that of direct-machined, conical feedhorns using the components depicted in Fig.~\ref{fig:850setup}.  
Alignment between the detectors and feedhorns is accomplished using a combination of alignment pins and visual fine tuning under a microscope. 
The machined precision of the direct-machined aluminum feedhorn locations was not sufficient to fully align all available detector-feedhorn pairs at the same time.  Therefore, only one feedhorn-detector set was optimally aligned and measured as shown in Fig.~\ref{fig:850meas}.

The detectors and feedhorns are coupled to a beam-filling and temperature-controlled cryogenic blackbody.
Four metal-mesh, low-pass filters \cite{ade} are used to control the radiation that couples to the detectors. 
Due to availability at the time of measurement, we used filters with a $\sim$1~THz low-pass cutoff rather than the final CCAT 850~GHz band target of 900~GHz. 
The waveguides in each prototype feedhorn structure were designed to provide a high-pass cutoff at 800~GHz. Therefore, the measured room-temperature transmission spectra of the metal-mesh filters are used to model the incident power for a given blackbody temperature in the 800--1000~GHz band.

The optical efficiency of the detector and feedhorn can be determined from measurements of the
noise equivalent power (NEP) as a function of incident optical power.
We model the noise as a combination of photon wave and shot noise, detector recombination noise and a detector/readout noise floor term as described in detail in \cite{vaskuri2025}.
At high optical powers, the detector noise becomes photon-noise limited and provides a strong constraint on the optical efficiency. 
We find fitted detector+feedhorn band-average efficiencies of $\eta_{\rm{opt,Si}} = 0.74\pm 0.03$ and $\eta_{\rm{opt,Al}} = 0.74\pm0.06$ for the silicon and aluminum feedhorns, respectively, with the listed errors representing the 1$\sigma$ statistical uncertainty in the fitted efficiency due solely to measurement variance.
We conservatively ascribe an additional $\sim$10\% systematic uncertainty to these measurements due to various sources of potential bias as described in \cite{vaskuri2025}.  For these measurements, this bias is dominated by uncertainty in the bandpass model of sourced light that ideally reaches the detector, i.e. uncertainty in the cryogenic transmission of the low-pass filters (common to both measurements) and residual uncertainty in the full-length waveguide diameters 
(unique to each feedhorn type).

The prototype detectors used in this measurement were not optimized for the lab-measurement band of 800--1000~GHz, nor do they utilize a choke structure between the waveguide and detector as will be the case for the final detector array.
3-dimensional electromagnetic simulations of this experimental configuration over the expanded measurement band suggest that, for lossless feedhorns with no reflections, we should expect detector optical efficiencies of $\eta_{\rm{opt,Si}}=0.76$ and $\eta_{\rm{opt,Al}}=0.73$, 
with the difference between feedhorn types resulting from slightly different measured waveguide aperture diameters and an estimated $\sim$10~$\mu$m difference in the waveguide-detector vacuum gap of each measurement setup.

The finite conductivity of the feedhorns at millimeter wavelengths will add additional conductor losses that affect the total efficiency.  
Assuming bulk gold conductivity and an effective surface roughness of 50~$\mu$m (inferred from measurements of similarly produced waveguides \cite{koc2026}) for the silicon prototype, and 6061-T6 Aluminum in the high-roughness limit for the Al feedhorns, we would expect band-averaged room-temperature conductor losses of $\sim$15\% and $\sim$10\% for the long silicon prototype and shorter aluminum feedhorns, respectively. 
It is reasonable to expect that the conductivities of the feedhorn surfaces may increase at low temperatures, until limited by the anomalous skin effect (ASE) \cite{pippard1947}.
Calculated estimates of the ASE-limited conductivity, as well as published measurements at submillimeter wavelengths (e.g. \cite{kamikura2010}), suggest the loss in the gold-plated structures may be reduced by a factor of $\sim$2.    
The cryogenic conductivity of the aluminum alloy feedhorns is less well known, but if we assume a similar reduction in loss when cold, both measurement setups would have an expected total optical efficiency of $\sim$70\%, which is consistent with the measurements.

\section{Discussion and Conclusions}

These devices represent the first silicon-platelet feedhorn arrays designed for operation fully in the submillimeter.  
The 350~GHz and 850~GHz devices represent the first produced kilopixel-scale arrays of submillimeter feedhorns, and are expected to be the first submillimeter feedhorns of this type to be deployed to an on-sky instrument. 
The measured beam properties of the 350~GHz feedhorn arrays are well matched to simulation and exhibit high levels of array uniformity.
 
The fabrication of multiple prototype 850~GHz feedhorn sub-arrays were successful with metallization and platelet alignment consistent with that achieved on previous, longer wavelength devices.  
Although directly measured beammaps of the 850~GHz feedhorns are not currently available, 
cryogenic measurements show the silicon 850~GHz feedhorns exhibit high levels of transmission with optical efficiency that is consistent with both simulation and that demonstrated with traditional, direct-machined aluminum feedhorns at millikelvin temperatures. 
The optical efficiency measurements are consistent with there being minimal ($\lesssim10$\%) cryogenic conductor loss and reflection in the silicon-platelet feedhorns. 

We note that the 850~GHz detectors, feedhorns and waveguide interfaces used in this demonstration were not optimized for the 800--1000~GHz band used for these measurements. The final optimized designs have an improved simulated efficiency of $\eta_{\rm{opt}}=0.94$ for lossless feedhorns within the target science band of 800--900~GHz.
The final deployment silicon feedhorn design, which is shorter with a larger diameter waveguide than the prototype, has simulated conductor losses of $\sim$5\% and $\sim$10\% at cryogenic and room temperatures, respectively, averaged across the science band.
Combining the detector and feedhorn efficiencies, the cryogenic 850~GHz focal plane is predicted to have a total per-channel optical efficiency of 89\%.

The first full 850~GHz array of over 6,000 feedhorns has been fabricated and will soon begin characterization when coupled to the first CCAT 850~GHz science-grade detector array.
Overall, the results presented here demonstrate the potential applicability of the technology to future submillimeter experiments and opens the possibility of pushing the technology to frequencies $\ge$1~THz for orbital or sub-orbital platforms (e.g. \cite{coppi2024blast}).

\section*{Acknowledgments and Notes}
We thank the anonymous reviewers for their helpful comments that helped lead to a significantly expanded and improved manuscript.

Certain commercial instruments and products are identified to specify the experimental study adequately. This does not imply endorsement by NIST or that the instruments and/or products are the best available for the purpose.

The 850~GHz feedhorn development was supported in part by the Canada Foundation for Innovation (CFI) and NSERC. 
350~GHz feedhorn development is supported by NSF award AST-2117631.
The CCAT project, FYST and Prime-Cam instrument have been supported by generous contributions from the Fred M. Young, Jr. Charitable Trust, Cornell University, and the CFI and the Provinces of Ontario, Alberta, and British Columbia. 
The construction of the FYST telescope was supported by the Großger{\"a}te-Programm of the German Science Foundation (Deutsche Forschungsgemeinschaft, DFG) under grant INST 216/733-1 FUGG, as well as funding from Universität zu Köln, Universität Bonn and the Max Planck Institut für Astrophysik, Garching.

\bibliographystyle{IEEEtran}
\bibliography{journals,jh,mkidbib,citations} 

\end{document}